\documentclass[showpacs,showkeys]{revtex4}
\usepackage{graphicx}

\usepackage{amsmath}
\usepackage{amssymb}
\usepackage{array}
\usepackage{color}

\begin{document}

\title{Asymptotically flat gravitating spinor field solutions. \\
Step 1 - the statement of the problem and \\
the comparison with confinement problem in QCD.}

\author{Vladimir Dzhunushaliev}
\affiliation{Department of Physics and Microelectronic
Engineering, Kyrgyz-Russian Slavic University, Bishkek, Kievskaya Str.
44, 720021, Kyrgyz Republic \\
and \\
Institute of Physics of National Academy of Science
Kyrgyz Republic, 265 a, Chui Street, Bishkek, 720071,  Kyrgyz Republic}
\email[Email: ]{vdzhunus@krsu.edu.kg}

\date{\today}

\begin{abstract}
The situation with asymptotically flat gravitating spinor field solutions is considered. It is supposed that the problem of constructing these solutions is connected with confinement problem in quantum chromodynamics. It is argued that in both cases we must use a nonperturbative quantum technique.
\end{abstract}


\maketitle

\section{The statement of the problem}

At the moment in general relativity  we know asymptotically flat solutions for all kind of mater (black holes with scalar, electric, magnetic and nonabelian gauge fields, as well as particle-like solutions with nonabelian gauge fields) with the exception of a spinor field. Most likely there were attempts to find such kind of solutions for the gravitating spinor field. For instance there are solutions describing the propagation of 1/2-spin particles on the background of Kerr solution (for details, see \cite{chandrasekhar}). One can suppose that after that there were the attempts to find self-consistent solutions of Einstein - Dirac equations. But we see that these efforts until now are unsuccessful. It is natural to ask: why asymptotically flat solutions in Einstein - Dirac gravity do not exist ?

We have two answers on this question only:
\begin{enumerate}
	\item The asymptotically flat spacetime for island spinor field distribution does  exist but the finding of the solutions runs into technical problems. That is the problem is purely technical one.
	\item There exists a fundamental physical problem connected with obtaining such kind solutions.
\end{enumerate}
In this paper we would like to argue in favor of the second possibility. For this we will compare the influence of gravity and SU(3) gauge fields on spinor fields. The Lagrangian for the spinor field interacting with gravity is
\begin{equation}
	\mathcal L_{\psi - grav} = \bar \psi \gamma^\mu \left(
		\partial_\mu - \frac{1}{4}
		\omega_\mu^{\phantom{\mu} \bar \mu \bar \nu}\gamma_{\bar \mu \bar \nu}
	\right) \psi
\label{10}
\end{equation}
where $\bar \mu=\bar 0, \bar 1, \bar 2, \bar 3$ is the Lorentz index;
$\bar \psi = \bar \psi^\dagger \gamma^{\bar 0}$ is the Dirac conjugated spinor;  $\mu=0,1,2,3$ is the world index; $e_{\bar \mu}^{\phantom{\bar \mu} \mu}$ is the tetrada; $\gamma^{\bar \mu}$ are the 4D Dirac matrices in a flat Minkowski space;
$\gamma_{\bar \mu \bar \nu} =
\frac{1}{2}\left(\gamma_{\bar \mu}\gamma_{\bar \nu} -
\gamma_{\bar \nu}\gamma_{\bar \mu} \right)$. All definitions for spinor differential geometry are taken from \cite{ortin}. The interaction between gravity and spinor field is described by the term
\begin{equation}
	\mathcal L_{\psi - grav - int} = - \frac{1}{4}
	\bar \psi \gamma^\mu \omega_\mu^{\phantom{\mu} \bar \mu \bar \nu}
  \gamma_{\bar \mu \bar \nu} \psi .
\label{15}
\end{equation}
The Lagrangian for the spinor field interacting with the SU(3) gauge field (for the Minkowski spacetime) is
\begin{equation}
	\mathcal L_{\psi - SU(3)} = \bar \psi \gamma^\mu \left(
		\partial_\mu - i g A^a_\mu t^a
	\right) \psi
\label{20}
\end{equation}
here $\mu = 0,1,2,3$ is the index in the Minkowski spacetime; $a=1,2, \cdots , 8$ is the color index; $A^a_\mu$ is the SU(3) gauge potential; $t^a = \lambda^a/2$ are the SU(3) generators; $\lambda^a$ are Gell-Mann matrixes; $[t^a, t^b] = i f^{abc} t^c$; $f^{abc}$ are the SU(3) structural constants. The interaction between SU(3) gauge field and spinor field is described by the term
\begin{equation}
	\mathcal L_{\psi - SU(3) - int} =
	- i g \bar \psi \gamma^\mu A^a_\mu t^a \psi .\label{25}
\end{equation}
It is well known that in quantum chromodynamics the strong interaction between the components $A^a_\mu$ of the gauge fields leads to a quantum non-perturbative effect: quark confinement or unobservability of a single quark. The definition of quark confinement one can find in Wikipedia \cite{wiki}: ``Color confinement is the physics phenomenon that color charged particles (such as quarks) cannot be isolated singularly, and therefore cannot be directly observed. The reasons for quark confinement are somewhat complicated; there is no analytic proof that quantum chromodynamics should be confining, but intuitively, confinement is due to the force-carrying gluons having color charge. As any two electrically-charged particles separate, the electric fields between them diminish quickly, allowing (for example) electrons to become unbound from nuclei. However, as two quarks separate, the gluon fields form narrow tubes (or strings) of color charge, which tend to bring the quarks together as though they were some kind of rubber band. This is quite different in behavior from electrical charge. Because of this behavior, the color force experienced by the quarks in the direction to hold them together, remains constant, regardless of their distance from each other.''

The interaction between gravity and spinor field \eqref{15} looks similarly to the term describing the interaction between SU(3) gauge and spinor fields \eqref{25}. Consequently if the spinor field creates a strong gravitational field then we have the full analogy with quantum chromodynamics. It allows us to say that in some situations the self-consistent Einstein - Dirac theory may lead to the situations that are analogous to some situations in quantum chromodynamics. We know that the confinement problem in quantum chromodynamics can not be resolved on the classical level: only a non-perturbative quantized SU(3) gauge theory may resolve the confinement problem.

Thus we can assume that a gravitating spinor field may create a physical phenomenon which can be described on a quantum non-perturbative level only. On this basis we suppose that \textcolor{blue}{\emph{asymptotically flat solutions for the gravitating spinor field do exist taking into account a non-perturbative quantized gravitational field}}. Exactly as a flux tube between quark and antiquark can be described taking into account non-perturbative quantized SU(3) gauge theory.

In which situations it can take place ?

\section{Discussion and conclusions}

Quantum field theoretical calculations of the evaporation of a black hole are based on perturbative calculations (using Feynman diagrams). But if the final stage of the evaporation lies in the strong gravity region then the calculations must be non-perturbative ones. In this case  an interesting possibility there exists: if above mentioned asymptotically flat solutions do exist then one can suppose that the solutions describe final stage of the  black hole evaporation. Interesting possibility in this case is that the black hole does not disappear but appears a static object filled with a non-perturbative matter.

Finally we would like to present some physical arguments why above mentioned asymptotically flat solutions for spinor field do not exist in classical gravity. We know that the classical spinor field does not exists in the Nature. Dirac equation has the physical sense as the equation describing a quantum particle only. In quantum mechanics observables are: eigenvalues of an operator and mean value of a physical quantity only. The energy - momentum tensor for the spinor field is
\begin{equation}
 T_{\bar \mu}^{\phantom{\bar \mu} \mu } = - \frac{i}{2} \bar \psi
 \Gamma^{\bar \nu} \left(
		e_{\bar \mu}^{\phantom{\bar \mu} \mu} e_{\bar \nu}^{\phantom{\bar \nu} \nu} +
		\eta_{\bar \mu \bar \nu} g^{\mu \nu}
	\right) D_\nu \psi +
	(\text{Hermitian conjugated})
\label{3-30}
\end{equation}
where
$D_\nu = \partial_\mu - \frac{1}{4}	
\omega_\mu^{\phantom{\mu} \bar \mu \bar \nu}\gamma_{\bar \mu \bar \nu}$ is the covariant derivative. We see that the energy - momentum tensor is neither an  eigenvalue nor a mean value. Consequently the RHS of Einstein equations is bad defined. The conclusion is : there exists some contradiction between gravity and quantum theory. The contradiction may be make oneself evident in the fact that in general relativity asymptotically flat solutions with spinor field do not exist.

Above mentioned reasonings allow us to offer a possible test for any quantum gravity theory: such theory should be able to describe asymptotically flat solutions with spinor field just as quantum chromodynamics should be able to describe a flux tube between quark and antuquark.

\section*{Acknowledgements}

I am grateful to the Research Group Linkage Programme of the Alexander von Humboldt Foundation for the support of this research.


\begin{thebibliography}{99}

\bibitem{chandrasekhar}
S. Chandrasekhar,
``The mathematical theory of black holes'',
Oxford University Press, 1983.

\bibitem{ortin}
T. Ortin,
``Gravity and Strings'',
Cambridge, UK ; New York : Cambridge University Press, 2004.

\bibitem{wiki}
Wikipedia,
http://en.wikipedia.org/wiki/Color$\_$confinement


\end{thebibliography}
\end{document}